\documentclass{elsart}
\usepackage{natbib}
\input{psfig}
\def\gsim{ \lower .75ex \hbox{$\sim$} \llap{\raise .27ex \hbox{$>$}} }
\def\lsim{ \lower .75ex\hbox{$\sim$} \llap{\raise .27ex \hbox{$<$}} }

\begin{document}

\runauthor{Ghisellini}
\begin{frontmatter}
\title{Extreme blazars}

\author[a]{Gabriele Ghisellini}
\address[a]{Osservatorio Astr. di Brera. Via Bianchi 46 Merate I-23807 Italy}

\begin{abstract}
TeV emission can be a common characteristic of low power blazars.
This is in line with the sequence of blazars relating
the observed bolometric luminosity with their overall
spectral energy distribution.
Detecting new TeV blazars, possibly at different redshifts, is
important for studying the particle acceleration process
operating in the jet, and it is even more important
for studying the far infrared--optical background.
Recent studies of low power BL Lacs suggest that
the particle acceleration mechanism may be a two--step process,
analogous to what is invoked to explain the spectra
of gamma--ray bursts.
I will briefly discuss these findings, together with a simple 
scheme for selecting good TeV candidates and I will briefly 
comment on the very fast TeV and X--ray variability observed in BL Lacs.
\end{abstract}
\end{frontmatter}

\section{Introduction}

\begin{figure}
\psfig{figure=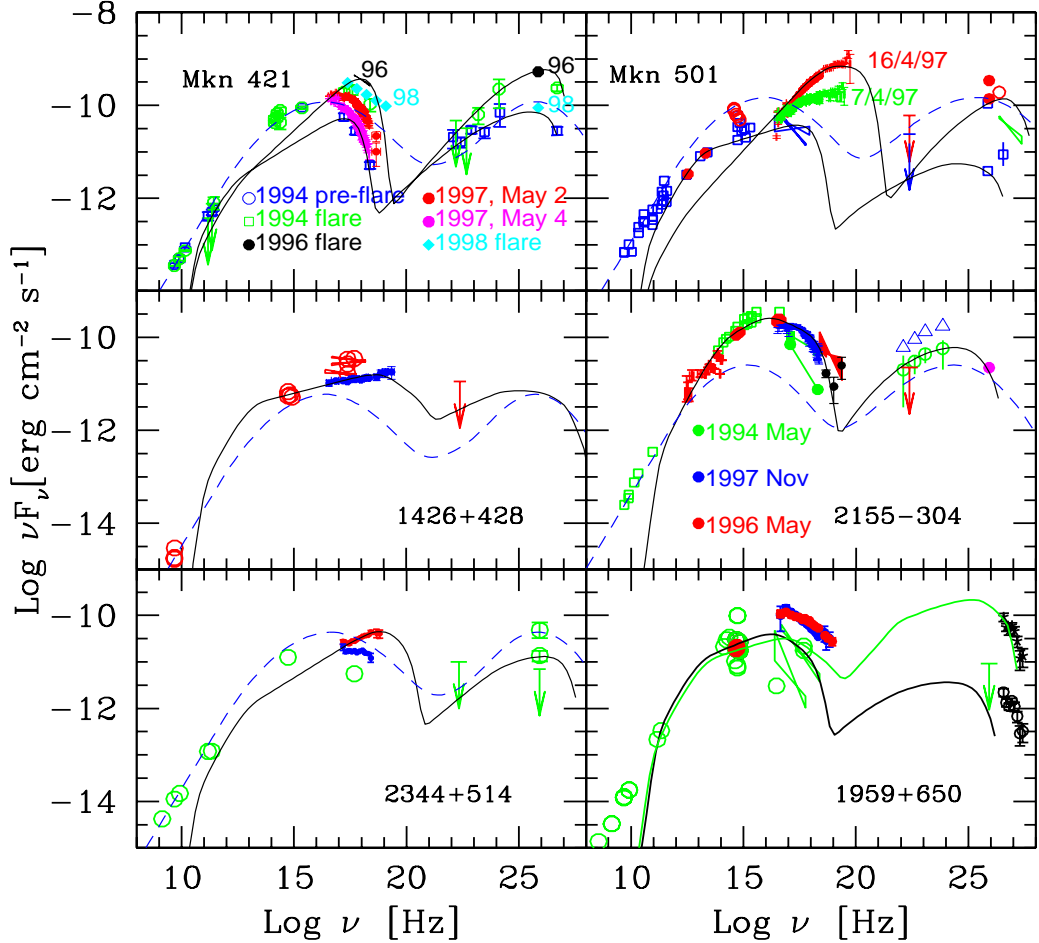,width=14cm,height=13.5cm}
\caption{The SED of the 6 BL Lacs detected so far at TeV
energies. The dashed and solid lines are model fits
of different states of the sources.
See Costamante \& Ghisellini (2002) for the
relevant sources of data and discussion of the models.}
\label{sed_tev}
\end{figure}

TeV astronomy is now leaving its infancy and becoming adult.
The few BL Lacs (see Fig. \ref{sed_tev})
detected by the existing Cherenkov telescopes
will become hundreds when the promised tenfold increased sensitivity
(and lower energy thresholds)
of the new generation of telescopes will be reached.
It is a new electromagnetic window which has been opened
to scrutiny: the fact that it can be done on ground
means that it is one order of magnitude (at least) less expensive
than satellites, and this in turn means fast progress.
There are at least two science pillars involving TeV
astronomy: the first is of course the study of TeV 
emitting objects, and the violent physics involved,
while the second is the cosmological issue
of estimating the intergalactic infrared, optical 
(and in future UV, when the detecting threshold
will reach the 10 GeV range) backgrounds,
through the study of the absorption imprinted in the 
intrinsic spectrum of the TeV emitting sources through
the photon--photon collision process (e.g. Stecker \& De Jager 1997).
For these issues it is now important to find good
candidates for TeV emission within a range of redshifts.
As discussed below, we think that the best candidates are
among low power BL Lacs, characterized by a spectral energy
distribution (SED) with the first peak (in a $\nu F_\nu$
plot) at X--ray frequencies.
We call them {\it extreme} blazars.
We argue that the best TeV emitters are extreme BL Lacs 
with relatively strong radio emission, which measures 
the amount of seed photons required for the inverse Compton
scattering process.

\begin{figure}
\psfig{figure=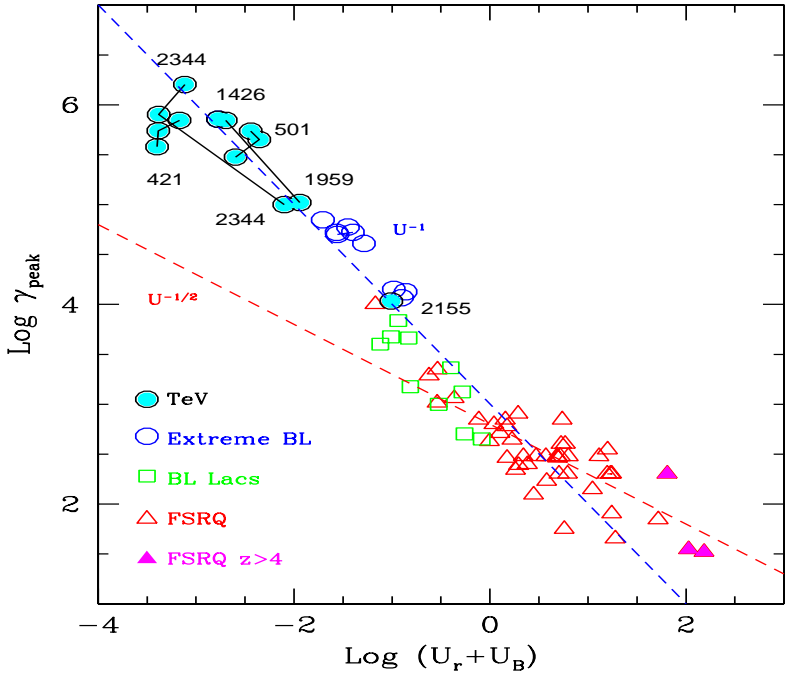,width=13cm,height=9.5cm}
\caption{The random Lorentz factor of the electrons emitting at the peaks of
the SED, $\gamma_{\rm peak}$, as a function of the comoving energy density
(radiative plus magnetic).
The points connected by a line correspond to the quiescent and flaring
state of the same source, as labeled.
The dashed lines correspond to $\gamma_{\rm peak} \propto U^{-1/2}$ and
$\propto U^{-1}$ (they are not best fits). From Ghisellini, Celotti \& Costamante (2002).
}
\label{gb_u}
\end{figure}

\section{Extreme blazars}

Besides the phenomenological approach leading to the blazar sequence 
discussed in Fossati et al. (1998) and in Donato et al. (2001),
to understand why blazars have their SED controlled by their bolometric
luminosity we need to model their spectra, in order to derive
the physical parameters of the emitting region such as
the size, the value of the magnetic field, the particle density and 
the beaming factor.
Ghisellini et al. (1998) considered the sample of blazars detected by 
EGRET with some spectral information in the EGRET band.
We applied a one--zone (leptonic) model, including (besides SSC)
the inverse Compton scattering process with photons produced 
externally to the jet.
We found a clear correlation between the energy of 
the electrons emitting at the peaks of the SED, $\gamma_{\rm peak}$, 
and the value of the magnetic plus radiation energy density 
measured in the comoving frame, $U$.
The correlation was of the kind $\gamma_{\rm peak} \propto U^{-1/2}$,
i.e. the radiative cooling rate at $\gamma_{\rm peak}$, 
$\dot\gamma\propto \gamma^2_{\rm peak}U$, is constant and the same for all sources.
Since the more extreme BL Lacs were under--represented in this work, 
Ghisellini, Celotti \& Costamante (2002) later extended 
the range of parameters 
by including sources with more extreme values of $\gamma_{\rm peak}$,
corresponding to low power BL Lacs.
This was possible through the {\it Beppo}SAX observations of
extreme blazars, covering the range 0.1--100 keV (i.e. 
all these BL Lacs were detected in the hard X--ray band).

We found a new branch of the correlation at high 
$\gamma_{\rm peak}$, with $\gamma_{\rm peak}\propto U^{-1}$.
We interpreted it as the effect of a finite timescale of injection, 
$t_{\rm inj}$:
in powerful blazars all particles cool in $t_{\rm inj}$ 
({\it fast cooling} regime),
while in low power BL Lacs the radiative cooling is less severe
({\it slow cooling} regime), and only the high energy particles 
can cool in this timescales.
If we take a snapshot of the spectrum at the end of the injection,
i.e. at the maximum of the flare, we then have two behaviors:
in the fast cooling regime $\gamma_{\rm peak}$
is determined by the low energy cutoff of the injected
distribution, since it is at this energy that 
a break in the final distribution occurs.
In the slow cooling regime, instead, $\gamma_{\rm peak}$ corresponds to
particles whose cooling time equals the injection time,
and we obtain $\gamma_{\rm peak}\propto U^{-1}$.

These results suggest a two--step acceleration process:
first a phase of ``pre--heating" determining
the low energy cut--off of the injected distribution, 
and  then a rapid acceleration leading to a 
non--thermal energy distribution.
This two--step process can operate in all blazars:
their different SED simply reflects the different degree
of radiative cooling.
The typical energies produced by the pre--heating 
(in the range $\gamma_{\rm min} \sim$10--10$^3$)
correspond to the balance of the heating and cooling rates:
this gives $\gamma_{\rm min}\propto U^{-0.5}$.
We have $\gamma_{\rm peak}=\gamma_{\rm min}$ in powerful blazars,
where electrons of all energies cool in a time $t_{\rm inj}$.
Instead, when the cooling is less severe, the particle
distribution (after $t_{\rm inj}$) will have a break at
$\gamma_{\rm peak}>\gamma_{\rm min}$, determined by  
$t_{\rm cool}(\gamma_{\rm peak})=t_{\rm inj}$, leading to
$\gamma_{\rm peak}\propto U^{-1}$.

Fig. \ref{gb_u} also shows that the parameters
corresponding to different states of specific TeV sources 
obey the general trend (see the ``tracks" connecting
different points belonging to the same source).
This is true even when the observed synchrotron peak frequency
increases with the observed luminosity (as during the 1997 flare 
of Mkn 501; Pian et al. 1998), contrary to the general
sequence discussed by Fossati et al. (1998).

\begin{figure}
\psfig{figure=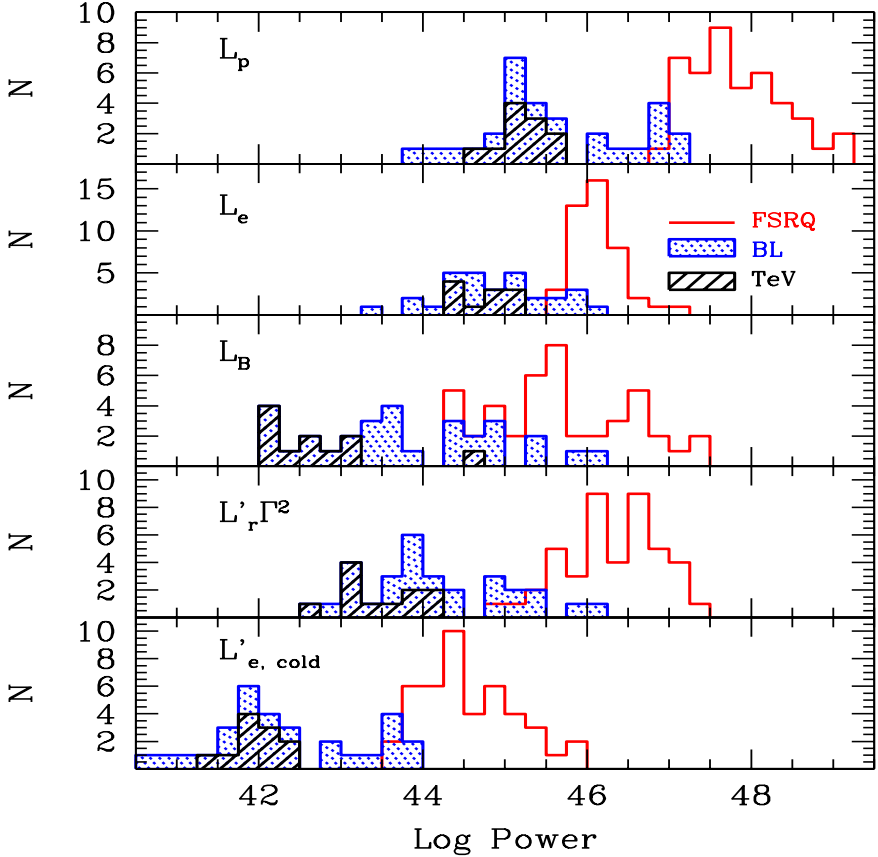,width=14cm,height=12cm}
\caption{Histograms of the distribution of kinetic power
(in erg s$^{-1}$)
of the blazars considered in Celotti \& Ghisellini 2003 (in prep.).
$L_{\rm p}$ is the power carried by protons assuming one
proton per emitting electron;
$L_{\rm e}$ is the power carried by the emitting electrons;
$L_{\rm B}$ is the Poynting flux;
$L^\prime_{\rm r}\Gamma^2$ is the power radiated by the jet
and $L_{\rm e, cold}$ is the power carried in the form of 
electron rest mass (i.e. the difference with
$L_{\rm e}$ is that the latter includes the relativistic random
energy of the particles).
The shaded and dotted histograms corresponds to TeV and
other BL Lac sources, as labeled.
}
\label{istopower}
\end{figure}

\subsection{Jet power of TeV sources}

The fact that the emission region producing most of the 
observed luminosity is well localized allows us
to estimate the total power carried by the jet.
In fact, through modeling, we derive the size,
the magnetic field, the particle density and
the bulk Lorentz factor of the emission region, and these
are the quantities needed to derive the power carried
in the form of bulk kinetic energy and in Poynting flux.
Fig. \ref{istopower} shows our results,
and we can compare TeV sources with the other blazars.
As can be seen, TeV sources have the same kinetic powers of 
other BL Lacs, but significantly less Poynting flux.
Contrary to more powerful blazars, we do not find 
strong constraints, on energetic grounds, to limit
the number of electron--positron pairs of their jets.
This is due to the fact that the mean electron
random Lorentz factor is very high in TeV sources, 
making the relativistic mass of electrons to be roughly equal
to the proton mass: it makes a little difference then 
to have an electron--proton or a pure pair plasma
(compare $L_{\rm e}$ with $L_{\rm p}$ in Fig. \ref{istopower}).

\begin{figure}
\psfig{figure=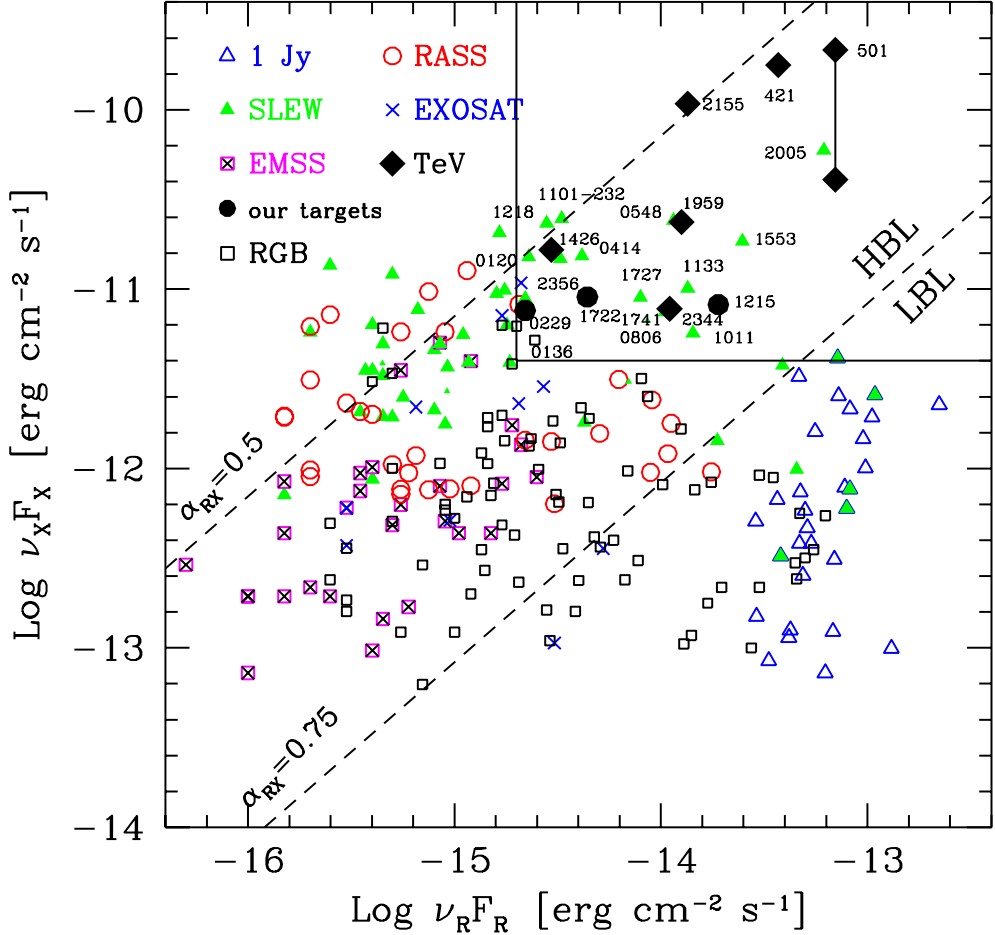,width=15cm,height=15cm}
\vskip -0.5 true cm
\caption{BL Lac objects in the radio (5 GHz) and X--ray (1 keV)
$\nu F_\nu$ plane.
Sources belonging to different samples have
different symbols, as labeled.
The objects marked with their name are good TeV candidates,
besides the objects already detected (filled circles).
}
\label{fxfr}
\end{figure}

\section{Candidates for strong TeV emission}

\begin{figure}
\psfig{figure=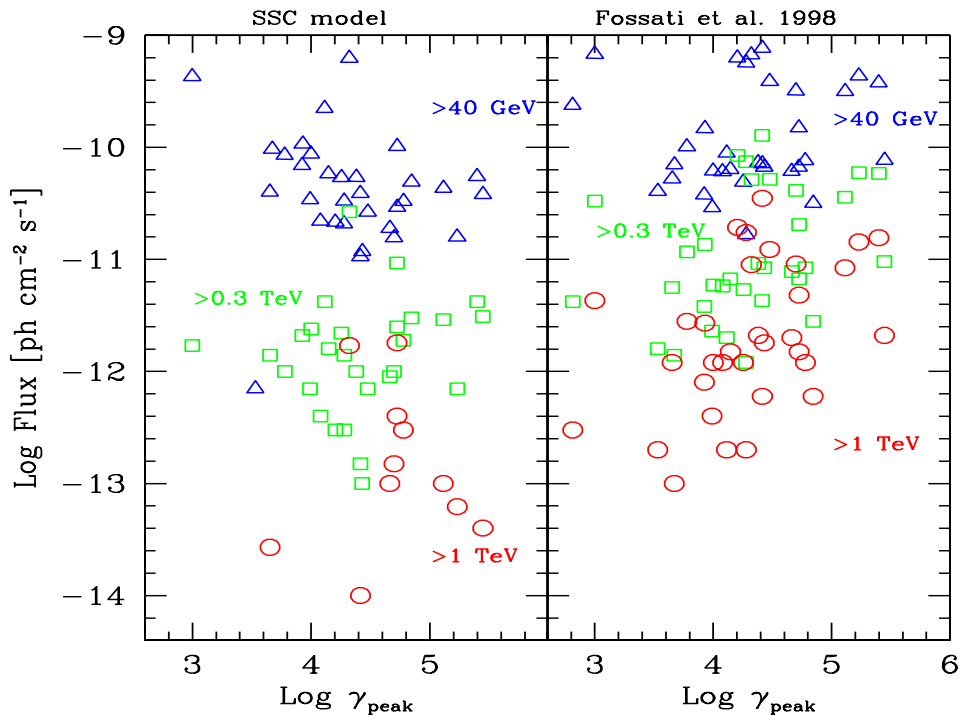,width=13.5cm,height=11cm}
\vskip -0.5 true cm
\caption{TeV fluxes predicted by 
a one--zone simple SSC model (left panel) and by the 
phenomenological model by Fossati et al. 1998,
as modified by Costamante \& Ghisellini 2002.
Note that the intergalactic absorption is not accounted for.
}
\label{predicted}
\end{figure}

To emit TeV photons we need TeV electrons, which emit, 
by the synchrotron process, in the X--ray band
(for typical magnetic fields of 0.1--1 Gauss and beaming factors
$\delta\sim 10$).
Therefore sources with a strong synchrotron X--ray flux
are obviously good TeV candidates.
On the other hand, to have a large TeV luminosity 
through synchrotron self Compton (SSC),
we need also synchrotron IR--optical seed photons, 
and this leads to a trade-off, since
sources with a prominent synchrotron peak at hard X--ray energies
have relatively little IR--optical emission.
Fig. \ref{fxfr} shows how BL Lac objects (belonging to several
samples) are located in the 
X--ray -- radio flux plane. 
Filled diamonds corresponds to the already TeV detected BL Lacs:
all of them are contained in the rectangle limited by the 
solid lines. 
We can see that all the TeV BL Lacs are HBL (high frequency peak
BL Lacs, Giommi \& Padovani 1994) characterized by a broad band radio
to X--ray spectral index $\alpha_{\rm RX} < 0.75$ (lower dashed line),
but all of them are relatively bright radio sources.
We believe that this occurs because the radio flux
is a measure of the amount of IR--optical emission,
providing the seed photons for the scattering.
It can also be seen that there are several other sources
in the vicinity of the already detected TeV ones.
We considered them as the best TeV candidates, and
we plot in Fig. \ref{predicted} the expected flux at three
different energy thresholds according to two models:
a simple SSC model and the phenomenological model
discussed in Fossati et al. (1998) with the modifications
introduced in Costamante \& Ghisellini (2002).

\section{Extreme variability}

Besides the famous example of very rapid variability in
the TeV band shown by Mkn 421 in 1996 (Gaidos et al. 1996)
there have been other examples of nearly factor 2 
flux variations in a timescale of $\sim$5--20 minutes: 
in Mkn 501 in X--rays (Catanese \& Sambruna 2000), 
BL Lac in X--rays (Ravasio et al. 2002)
and Mkn 421 again in the TeV band (Aharonian et al. 2002).
This very rapid variability is however not common, as
indicated by the power spectra and structure functions
(e.g. Fossati et al. 2000; Tanihata et al. 2003) 
calculated for the same sources, which suggests a 
more relaxed typical timescale of order of a few hours.
The ultra--fast variations must be produced by very compact
regions of size $\sim10^{14}$ cm,
several times smaller than the ``normal" emission regions, 
and one interesting possibility is that they may correspond
to ``explosive" events occurring {\it within} larger and
relatively steadier emission regions.
Just for illustration, consider two colliding shells
(as in the internal shock scenario),
dissipating energy throughout their volumes, but also 
triggering reconnection of the 
magnetic field lines in small sub--volumes, 
dissipating large amounts of energy in the form of 
relativistic particles.
Besides their own synchrotron flux, these particles 
will also scatter the ``ambient" photons produced by the
larger region, enhancing the inverse Compton contribution.
If this idea is true, then we expect that these fast
variations of the inverse Compton flux occur on top of flares
(produced by the shell--shell collision event), 
and not during quiescent states (Ghisellini \& Tavecchio, in prep.).

\section{Internal shock scenario}

The complex phenomenology of blazars and of blazar jets
can be framed in a coherent picture, in which the central
engine does not work continuously, but is injecting
power in the jet intermittently, through shells
of matter moving at slightly different velocities.
These shells catch up at some distance from the black hole, 
and collide dissipating part of their kinetic energy.
This ``internal shock" scenario is one of the
leading ideas for the generation of the prompt emission 
of gamma--ray bursts, and can work even better for blazar jets.
As in gamma--ray bursts, the typical variability timescales 
in this scenario is linked with the initial distance $\Delta R$ between
two consecutive blobs, which should be of order of
the Schwarzchild radius (they collide at a distance $\sim\Gamma^2\Delta R$,
but the observed time variability is shortened by a factor $\Gamma^2$
by the Doppler effect).
This scenario is easy to visualize, it explains
the main characteristics of blazar jets and it allows
numerical estimates
(e.g. Ghisellini 1999; Spada et al. 2001; Tanihata et al. 2003).


\begin{thebibliography}{999}

\bibitem{a} Aharonian F. et al., 2002, A\&A, 393, 89
\bibitem{cs} Catanese M. \& Sambruna R.M., 2000, ApJ, 534, L39
\bibitem{gc} Costamante L. \& Ghisellini G., 2002, A\&A
\bibitem{donato} Donato D. et al. 2001, A\&A, 375, 739 
\bibitem{fmccg} Fossati G. et al. 1998, MNRAS, 299, 433 
\bibitem{f} Fossati G. et al., 2000, ApJ, 541, 153 
\bibitem{ga} Gaidos J.A. et al., 1996, Nature, 383, 319
\bibitem{g} Ghisellini G., 1999, Astronomische Nachrichten, 320, p. 232
\bibitem{gcfmc} Ghisellini G. et al., 1998, MNRAS, 301, 451 
\bibitem{gcc} Ghisellini G., Celotti A. \& Costamante L., 2002, A\&A
\bibitem{gp} Giommi P. \& Padovani P., 1994, ApJ, 444, 567
\bibitem{pian} Pian et al. 1998, ApJ, 491, L17  
\bibitem{r} Ravasio M. et al., 2002, A\&A, 383, 763 
\bibitem{s} Spada M. et al., 2001, MNRAS, 325, 1559
\bibitem{sdj} Stecker F.W. \& De Jager O.C., 1997, ApJ, 476, 712
\bibitem{t} Tanihata C. et al., 2003, ApJ, 584, 153


\end{thebibliography}
\end{document}